\documentclass[a4paper]{jpconf}
\usepackage{graphicx}
\usepackage{mathtools}

\begin{document}
\title{Prospect for top quark FCNC searches at the FCC-hh}

\author{Petr Mandrik on behalf of the FCC study group}

\address{NRC ``Kurchatov Institute'' – IHEP, Protvino}

\ead{Petr.Mandrik@ihep.ru}

\begin{abstract}
FCC-hh is a proposed future energy-frontier hadron collider, which goal is to provide high luminosity proton collisions at a centre-of-mass energy of 100 TeV. 
The FCC-hh has an extremely rich physics program ranging from standard model (SM) measurements to direct searches for physics beyond the standard model (BSM).
One of the processes sensitive to new physics is flavour-changing neutral currents (FCNC) that extremely rare in the SM but have enhanced behavior in several BSM scenarios. 
In this report we present results of projections of FCNC searches in top quark interactions to the FCC-hh conditions based on Monte-Carlo simulation of FCC-hh detector.
\end{abstract}

\section{Introduction}
\begin{figure}
  \centering
  \includegraphics[width=0.9\linewidth,clip]{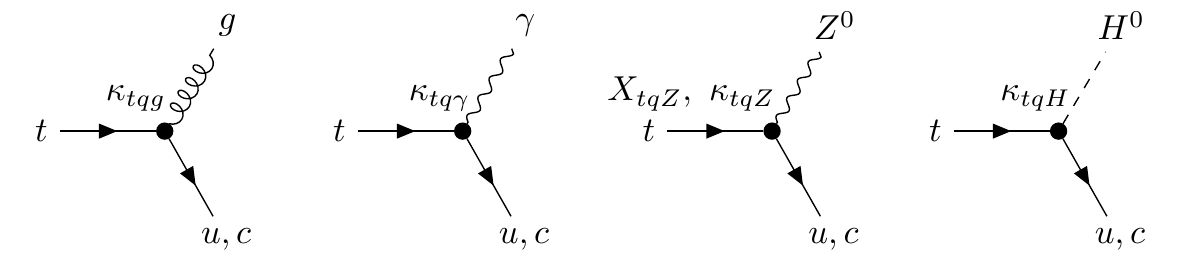}
  \iffalse
  \begin{subfigure}[t]{0.24\textwidth}
    \centering
    \includegraphics[width=\linewidth,clip]{fey_1.pdf}
  \end{subfigure}
  \begin{subfigure}[t]{0.24\textwidth}
    \centering
    \includegraphics[width=\linewidth,clip]{fey_2.pdf}
  \end{subfigure}
  \begin{subfigure}[t]{0.24\textwidth}
    \centering
    \includegraphics[width=\linewidth,clip]{fey_3.pdf}
  \end{subfigure}
  \begin{subfigure}[t]{0.24\textwidth}
    \centering
    \includegraphics[width=\linewidth,clip]{fey_4.pdf}
  \end{subfigure}
  \fi
  \caption{Diagrams for top quark decays mediated by FCNC couplings.}
  \label{fcnc_decays}       % Give a unique label
\end{figure}

The FCC-hh project, defined by the target of 100 TeV proton-proton collisions with a total integrated luminosity of 30 ab$^{-1}$,
will allow to extend the searches for flavour-changing neutral currents (FCNC, figure \ref{fcnc_decays})
forbidden in Standard Model (SM) at tree level and are strongly suppressed in loop corrections by the Glashow-Iliopoulos-Maiani mechanism \cite{PhysRevD.2.1285}.
The predicted SM branching fractions for top quark FCNC decays are expected to be $\mathcal{O}(10^{-12} - 10^{-17})$ \cite{Agashe:2013hma}
and are not expected to be detectable at the FCC-hh experimental sensitivity.
However, certain scenarios beyond the SM (BSM), such as
two-Higgs doublet model, warped extra dimensions and minimal supersymmetric models, 
incorporate significantly enhanced FCNC behavior that can be directly probed at the future collider experiments \cite{Agashe:2013hma}.
Observation of such processes would be a clear signal of new physics.

FCNC searches in top quark sector are typically based on the selection of events with isolated, well separated objects. 
On the other hand due to the expected increase of the energy of future collider experiments a significant number of events will contain high-energetic, 
boosted objects that require an exploration of different analysis strategy. 
We study the sensitivity of the FCC-hh to $t \rightarrow q\gamma$ and $t \rightarrow qH$ FCNC
transitions using the $pp \rightarrow t\bar{t} \rightarrow tq\gamma$ 
and $pp \rightarrow t\bar{t} \rightarrow tqH$ processes respectly where $q$ is a $u$ or $c$ quark.
The analyzes exploit the boosted regime where top-quark $p_T$ is much larger than its mass.
The signature of the signal processes includes high transverse momentum t-jet and a fat jet
clustered from collinear photon or Higgs decay products and light-flavour jet.
Resolved analysis of the FCNC in $tq\gamma$ via single top production in association with photon is described in \cite{Oyulmaz:2018irs}.
In \cite{Papaefstathiou:2017xuv} study of the FCNC in $tqH$ has covered the $H \rightarrow \gamma \gamma$ decay.
In this analyses the dominant Higgs decay channel $H \rightarrow b \bar{b}$ is explored. The study is based on ``fast'' simulation of the ``reference'' FCC-hh detector \cite{Zaborowska:2018origin, Zaborowska:2018qxe, Faltova:2018ayl}.

\section{Monte Carlo samples}
While the flavor-violating couplings of the top may arise from different sources, for the signal simulation
the effects of BSM physics in top interactions
described by an effective field theory approach.
The most general effective Lagrangian can be written as \cite{AguilarSaavedra:2004wm} (terms up to dimension five):
\begin{align}
  \begin{aligned}
    -\mathcal{L} & = g_s \kappa_{tqg} \bar{q} 
                     (g_L P_L + g_R P_R)
                     \frac{i \sigma_{\mu\nu} q^\nu}{\Lambda} T^a t G^{a\mu}
                   + e \kappa_{tq\gamma} \bar{q} 
                     (\gamma_L P_L + \gamma_R P_R)
                     \frac{i \sigma_{\mu\nu} q^\nu}{\Lambda} t A^{\mu} + \\
                 & + \frac{g}{2 c_W} X_{tqZ} \bar{q} 
                     (x_L P_L + x_R P_R)
                     t Z^{\mu}
                   + \frac{g}{2 c_W} \kappa_{tqZ} \bar{q} 
                     (z_L P_L + z_R P_R)
                     \frac{i \sigma_{\mu\nu} q^\nu}{\Lambda} t Z^{\mu} + \\
                 & + \frac{g}{2\sqrt{2}} \kappa_{tqH} \bar{q}
                     (h_L P_L + h_R P_R)
                     t H^{\mu} + h.c.,
  \end{aligned}
\end{align}
where $P_L$ and $P_R$ are chirality projectors in spin space, $\kappa_{tqX}$ and $X_{tqZ}$ are effective couplings
for the corresponding vertices, $\Lambda$ is the scale of new physics. 

The following background processes are considered for the $tq\gamma$ signal: 
QCD $\gamma+$jets, $t\bar{t}$, $t\bar{t}+\gamma$, $W+jets$, $Z+jets$, single top production and single top in association with photon.
The following background processes are considered for the $tqH$ signal: 
QCD multijets, $t\bar{t}$ ($+W$, $+Z$, $+H$), $W+jets$, $Z+jets$ and single top production.

All signals and backgrounds are generated at leading order
using the {\scshape MG5\_}a{\scshape MC@NLO}~2.5.2~\cite{Alwall:2011uj} package, with subsequent showering and hadronization in {\scshape Pythia}~8.230~\cite{Sjostrand:2014zea}.
The detector simulation has been performed with the fast simulation tool {\scshape Delphes}~3.4.2~\cite{deFavereau:2013fsa} using the reference FCC-hh detector parametrisation.
No additional proton-proton collisions during a single bunch crossing is assumed in the simulation.
In order to take into account higher order QCD corrections K-factors are applied to the signals and background samples.

\section{Event selection and signal extraction}
Events of the $tq\gamma$ signal are selected by requiring exactly one photon with $p_T > 200$ GeV,
at least two jets with cone $R=0.4$ and $p_T > 30$ GeV (one of which must be $b$-tagged), at least two jets with cone $R=0.8$ (``fat'' jets) and $p_T > 30$ GeV and
one or zero leptons ($e$ or $\mu$) with $p_T > 25$ GeV.
The $\Delta R$ between selected photon and b-tagged jet should be greater than $0.8$.
The fat jets matching photon and $b$-tagged jet respectively are required to have $p_T > 400$ GeV.
All objects must have $|\eta| < 3$.

Events of the $tqH$ signal are selected by requiring at least one jet with cone $R=0.8$
with at least two b-tagged subjets (with cone $R=0.4$) which corresponds to the FCNC decay of top quasrk (FCNC fat jet) and at least one additional fat jet 
with b-tagged subjet which corresponds to the SM decay of top (SM fat jet).
The leading (subleading) selected fat jet should have $p_T > 500$ ($p_T > 300$) GeV.
The $\Delta \phi$ between selected leading fat jet and subleading fat jet should be greater than $1.0$.
All objects must have $|\eta| < 3$.
The subjets with cone $R=0.2$ from selected fatjets are used to form the Higgs and W boson candidates.

A Boosted Decision Tree (BDT) constructed within the TMVA framework \cite{TMVA2007}
is used to separate the signal signature from the background contributions. $10\%$ of events selected for training and the remainder are used in the statistical analysis
of the BDT discriminants with the CombinedLimit package.
For each background a 30\% normalisation uncertainty is assumed and incorporated in statistical model as nuisance parameter.
The asymptotic frequentist formulae \cite{Cowan:2010js} is used to obtain an expected upper limit on signal cross section based on an Asimov data set of background-only model.

The following input variables are used for the $tq\gamma$ signal:
$\tau_{21}$ variable \cite{Thaler:2010tr} of the fat jet matched to the photon ($\gamma$-jet),
$\tau_{21}$ and $\tau_{32}$ variables of b-tagged fat jet (b-jet),
masses of soft-dropped \cite{Larkoski:2014wba} $\gamma$-jet and b-jet,
$p_T$ of the photon, $\gamma$-jet and b-jet,
scalar product of the photon and $\gamma$-jet four-vectors,
scalar product of b-jet and $\gamma$-jet four-vectors and
masses of two soft-dropped fat jets most corresponds to the mass of top quark.

The following input variables are used for the $tqH$ signal:
soft-dropped masses, $p_T$, $\tau_{21}$, $\tau_{31}$, $\tau_{32}$ variables \cite{Thaler:2010tr} and scalar product of the selected fat jets, 
$p_T$ and masses of the Higgs from leading FCNC fat jet and W boson from leading SM fat jet, scalar product of the Higgs (W boson) candidate and corresponding fat jet,
masses of the Higgs candidate from leading SM fat jet and W boson candidate from leading FCNC fat jet,
and mass disbalance, defined as $|m^{SM}_{fat jet} - m^{FCNC}_{fat jet}| / \max{(m^{SM}_{fat jet}, m^{FCNC}_{fat jet})}$.

\begin{figure}
  \centering
  \includegraphics[width=0.49\columnwidth]{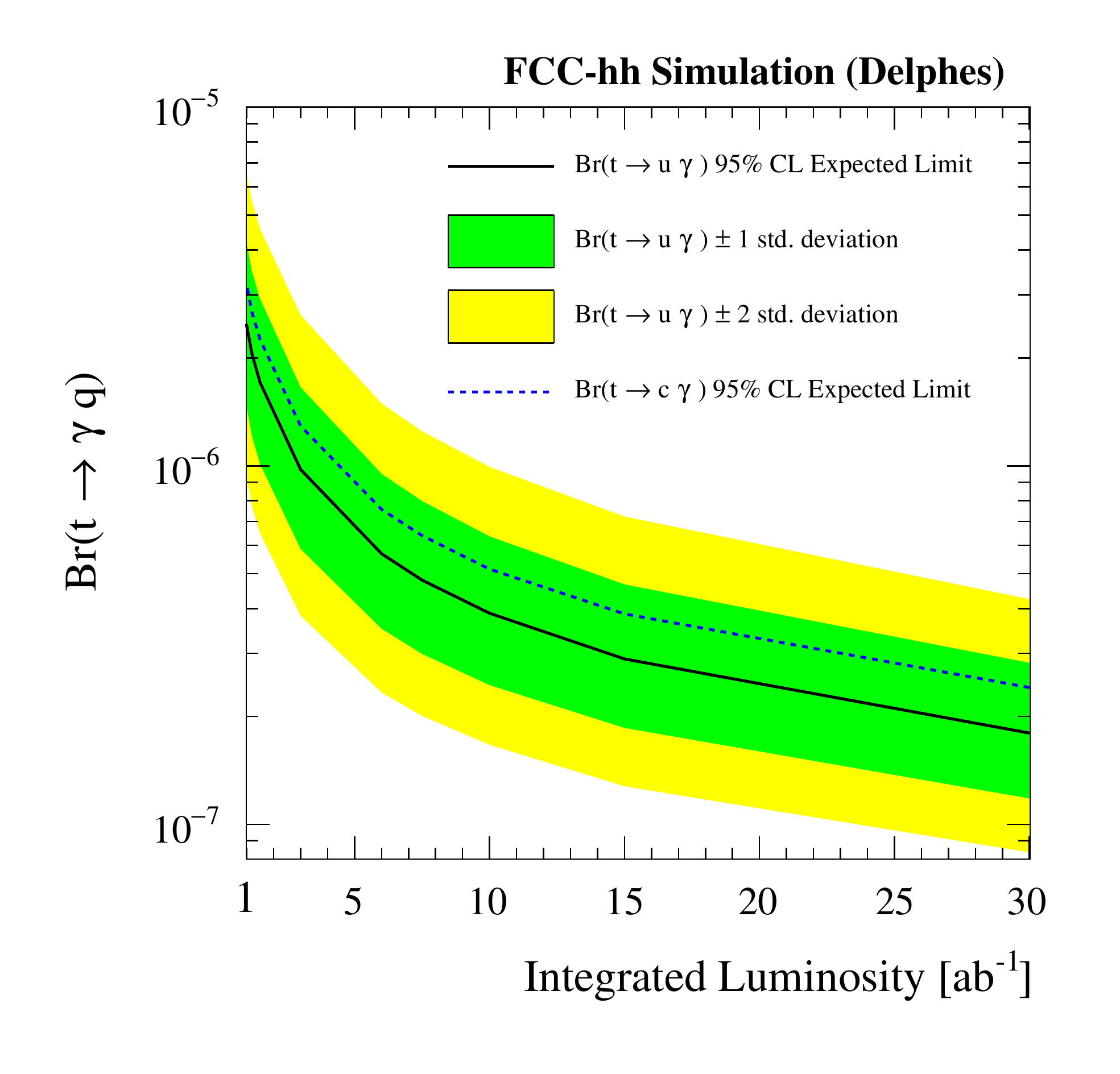}
  \includegraphics[width=0.49\columnwidth]{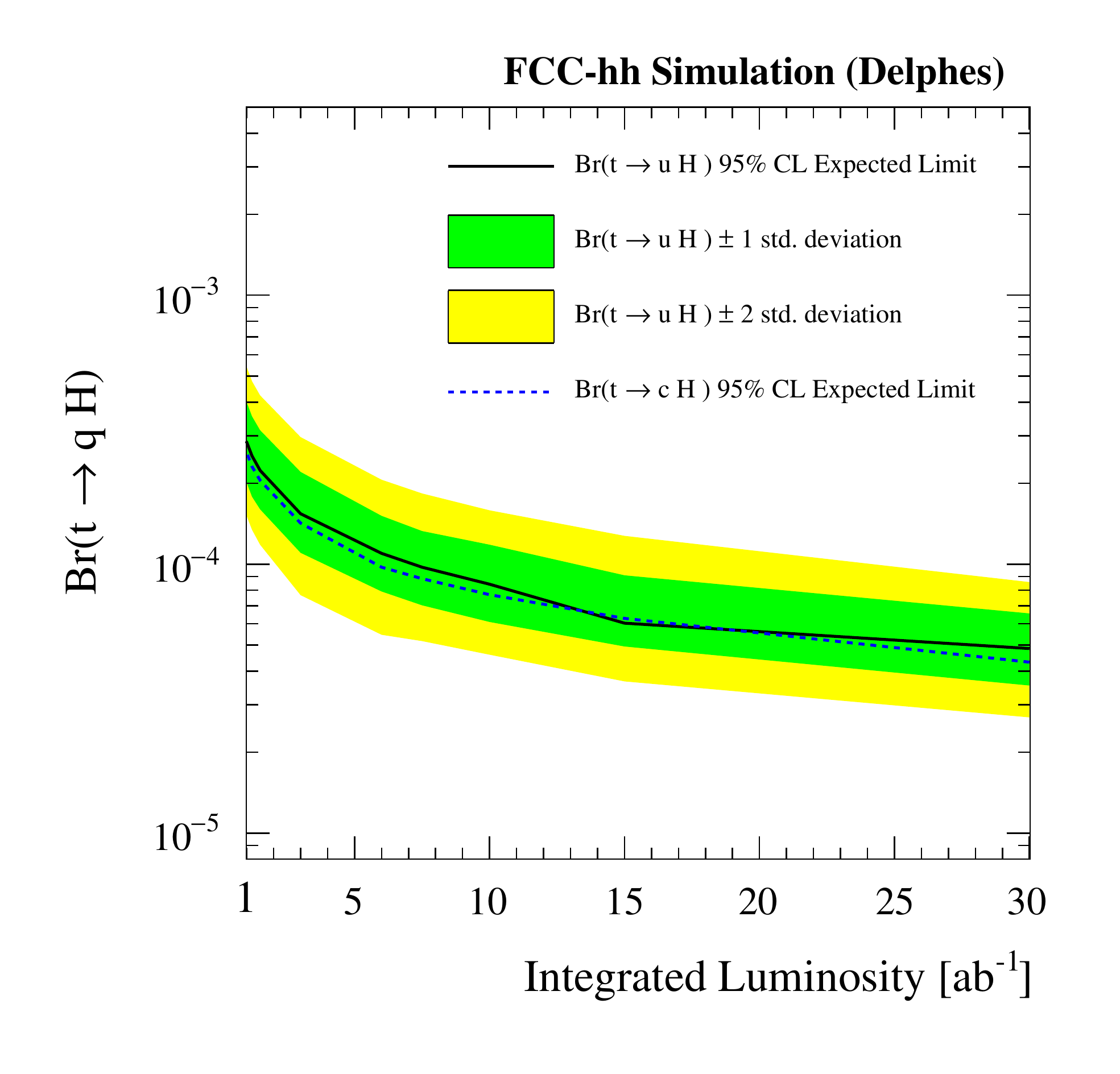}
  \caption{ Expected exclusion limits at 95\% C.L. on the FCNC $t \rightarrow q\gamma$ (left) and $t \rightarrow qH$ (right) branching fractions as a function of integrated luminosity. }
  \label{tqgamma_limits}
\end{figure}

    \begin{table}[h]
      \centering
      \label{tqgamma_limits_table}
      \caption{ The 95\% C.L. expected exclusion limits on the branching fractions for integrated luminosities of 30 ab$^{-1}$ and 3 ab$^{-1}$
                in comparison with present experimental limits and estimation for the HL-LHC.}
      \vspace{5pt}
      \renewcommand{\arraystretch}{1.5}
      \begin{tabular}{c|c c r}
      \hline
      \hline
      Detector                              & $\mathcal{B}(t \rightarrow u\gamma)$   & $\mathcal{B}(t \rightarrow c\gamma)$   & Ref. \\ \hline
      CMS (19.8 fb$^{-1}$, 8 TeV)           & $13  \times 10^{-5}$                   & $170  \times 10^{-5} $                 & \cite{Khachatryan:2015att} \\
      CMS Phase-2 (300 fb$^{-1}$, 14 TeV)   & $2.1 \times 10^{-5}$                   & $15   \times 10^{-5}$                  & \cite{Mandrik:2018gud} \\
      CMS Phase-2 (3 ab$^{-1}$, 14 TeV)     & $0.9 \times 10^{-5}$                   & $7.4  \times 10^{-5} $                 & \cite{Mandrik:2018gud} \\
      FCC-hh (3 ab$^{-1}$, 100 TeV)         & $9.8 \times 10^{-7}$                   & $12.9 \times 10^{-7}$                  & \\
      FCC-hh (30 ab$^{-1}$, 100 TeV)        & $1.8 \times 10^{-7}$                   & $2.4  \times 10^{-7} $                 & \\ \hline
      Detector                              & $\mathcal{B}(t \rightarrow uH)$        & $\mathcal{B}(t \rightarrow cH)$        & Ref. \\ \hline
      CMS (36.1 fb$^{-1}$, 13 TeV)          & $4.7 \times 10^{-3}$                   & $4.7 \times 10^{-3}$                   & \cite{Sirunyan:2017uae} \\
      ATLAS (36.1 fb$^{-1}$, 13 TeV)        & $1.9 \times 10^{-3}$                   & $1.6 \times 10^{-3} $                  & \cite{Aaboud:2018pob} \\
      FCC-hh (3 ab$^{-1}$, 100 TeV)         & $8.4 \times 10^{-5}$                     & $7.7 \times 10^{-5}$                 & \\
      FCC-hh (30 ab$^{-1}$, 100 TeV)        & $4.8 \times 10^{-5}$                     & $4.3 \times 10^{-5}$                 & \\
      \hline
      % 1.37442e-05 0.018821 9.81727e-06 0.0159067
      \end{tabular}
    \end{table}
    
\section{Results and conclusions}
To avoid ambiguities due to different normalizations of the couplings in the Lagrangian,
the branching ratios of the corresponding FCNC processes are used for presentation of the results.

The 95\% C.L. expected exclusion limits on the branching fractions are given in Table~\ref{tqgamma_limits_table}.
Figure~\ref{tqgamma_limits} shows the expected exclusion limits on the FCNC branching fractions as a function of integrated luminosity. 
This would improve the existing experimental limits \cite{Khachatryan:2015att} on the $t \rightarrow q \gamma$ branching fractions by about three-four orders of magnitude.
The limits on $\mathcal{B}(t \rightarrow cH)$, $\mathcal{B}(t \rightarrow uH)$ are comparable with the estimates of the limits on $\mathcal{B}(t \rightarrow qH)$ from \cite{Papaefstathiou:2017xuv}.
Further improvements can be obtained from the combinations with different analysis strategies such as resolved analysis and FCNC in production of the single top quark events.

% \begin{Table}
% \begin{center}
% \begin{tabular}{ c | c | c }
%   \hline \hline \rule{0pt}{3ex}
%   Process & Branching fraction for 30 ab$^{-1}$ (3 ab$^{-1}$) & Coupling strengths $\lambda$ for 30 ab$^{-1}$ (3 ab$^{-1}$) \\ \hline \rule{0pt}{3ex}
%   %$t\rightarrow u\gamma$ &  $1.44 \cdot 10^{-6}$  & $1.70 \cdot 10^{-3}$ \\
%   %$t\rightarrow c\gamma$ &  $1.96 \cdot 10^{-6}$  & $2.14 \cdot 10^{-3}$ \\
%   $t\rightarrow u\gamma$ &  $1.8 \cdot 10^{-7}$ $(9.8 \cdot 10^{-7})$ & $6.5 \cdot 10^{-4}$ $(15.1 \cdot 10^{-4})$ \\
%   $t\rightarrow c\gamma$ &  $2.4 \cdot 10^{-7}$ $(12.9 \cdot 10^{-7})$& $7.5 \cdot 10^{-4}$ $(17.3 \cdot 10^{-4})$ \\
%   %9.76375e-07 0.00151038 1.29069e-06 0.00173656
%   \hline \hline
% \end{tabular}
%   \caption{\label{tqgamma_limits_table} The 95\% C.L. expected exclusion limits on the branching fractions and coupling strengths $\lambda$ \cite{AguilarSaavedra:2004wm}
% for integrated luminosities of 30 ab$^{-1}$ and 3 ab$^{-1}$.}
%   \end{center}
% \end{Table}

\section*{Acknowledgments}
I would like to thank H.~Gray, C.~Helsens and S.~Slabospitskii for useful discussions.

\section*{References}
%\begin{thebibliography}{9}
%\bibitem{iopartnum} IOP Publishing is to grateful Mark A Caprio, Center for Theoretical Physics, Yale University, for permission to include the {\tt iopart-num} \BibTeX package (version 2.0, December 21, 2006) with  this documentation. Updates and new releases of {\tt iopart-num} can be found on \verb"www.ctan.org" (CTAN). 
%\end{thebibliography}
\bibliography{iopart-num}

\end{document}